\documentclass[aps,prl,reprint, nofootinbib,aas_macros,superscriptaddress]{revtex4-2}

\usepackage[utf8]{inputenc}
\usepackage{graphicx}
\usepackage{hyperref}
\hypersetup{colorlinks,bookmarksnumbered,
citecolor=blue, urlcolor=blue, linkcolor=blue}

\usepackage{amsmath}

\def\Planck{\textit{Planck}}

\usepackage[usenames, dvipsnames]{xcolor}

\definecolor{darkgreen}{RGB}{0,120,0}

\begin{document} 

\title{The Atacama Cosmology Telescope: A Test of the Gravitational Force Law on Cosmological Scales Using the Kinematic Sunyaev-Zeldovich Effect}

\author{P.~A.~Gallardo}
\affiliation{Department of Physics and Astronomy, University of Pennsylvania, 209 South 33rd Street, Philadelphia, PA, USA 19104}
\affiliation{Kavli Institute for Cosmological Physics, University of Chicago, Chicago, IL, 60637, USA}
\author{K.~Pardo}
\affiliation{Department of Physics \& Astronomy, University of Southern California, Los Angeles, CA, 90089, USA}
\author{O.~H.~E.~Philcox}
\affiliation{Department of Physics, Columbia University, New York, NY, 10027, USA}
\affiliation{Department of Physics, Stanford University, Stanford, CA 94305, USA}
\affiliation{Simons Society of Fellows, Simons Foundation, New York, NY 10010, USA}
\author{N.~Battaglia}
\affiliation{Department of Astronomy,  Cornell University, Ithaca, NY 14850, USA}
\author{E.~S.~Battistelli}
\affiliation{Sapienza University of Rome, Physics Department, Piazzale Aldo Moro 5, I-00185 Rome, Italy}
\author{R.~Bean}
\affiliation{Department of Astronomy,  Cornell University, Ithaca, NY 14850, USA}
\author{E.~Calabrese}
\affiliation{School of Physics and Astronomy, Cardiff University, The Parade, Cardiff, Wales CF24 3AA, UK}
\author{S.~K.~Choi}
\affiliation{Department of Physics and Astronomy, University of California, Riverside, CA 92521, USA}
\author{R.~D\"unner}
\affiliation{Instituto de Astrof\'isica, Facultad de F\'isica, Pontificia Universidad Cat\'olica de Chile, Avenida Vicu\~na Mackenna 4860, 7820436 Macul, Santiago, Chile}
\affiliation{Centro de Astro-Ingenier\'ia, Pontificia Universidad Cat\'olica de Chile, Av. Vicu\~na Mackenna 4860, 7820436 Macul, Santiago, Chile}
\author{M.~Devlin}
\affiliation{Department of Physics and Astronomy, University of Pennsylvania, 209 South 33rd Street, Philadelphia, PA, USA 19104}
\author{J.~Dunkley}
\affiliation{Joseph Henry Laboratories of Physics, Jadwin Hall, Princeton University, Princeton, NJ 08544, USA}
\affiliation{Department of Astrophysical Sciences, Peyton Hall, Princeton University, Princeton, NJ 08544, USA}
\author{S.~Ferraro}
\affiliation{Lawrence Berkeley National Laboratory, One Cyclotron Road, Berkeley, CA 94720, USA}
\affiliation{Berkeley Center for Cosmological Physics, Department of Physics, University of California, Berkeley, CA 94720, USA}
\author{Y.~Guan}
\affiliation{Dunlap Institute for Astronomy \& Astrophysics, University of Toronto, 50 St. George St., Toronto ON M5S 3H4, Canada}
\author{E.~Healy}
\affiliation{Kavli Institute for Cosmological Physics, University of Chicago, Chicago, IL, 60637, USA}
\author{C.~Herv\'ias-Caimapo}
\affiliation{Instituto de Astrof\'isica, Facultad de F\'isica, Pontificia Universidad Cat\'olica de Chile, Av. Vicu\~na Mackenna 4860, 7820436 Macul, Santiago, Chile}
\affiliation{Centro de Astro-Ingenier\'ia, Pontificia Universidad Cat\'olica de Chile, Av. Vicu\~na Mackenna 4860, 7820436 Macul, Santiago, Chile}
\author{M.~Hilton}
\affiliation{Wits Centre for Astrophysics, School of Physics, University of the Witwatersrand, Private Bag 3, 2050, Johannesburg, South Africa}
\affiliation{Astrophysics Research Centre, University of KwaZulu-Natal, Westville Campus, Durban 4041, South Africa}
\author{A.~D.~Hincks}
\affiliation{David A. Dunlap Department of Astronomy and Astrophysics, University of Toronto, 50 St. George St., Toronto ON M5S 3H4, Canada}
\affiliation{Specola Vaticana (Vatican Observatory), V-00120 Vatican City State}
\author{J.~C.~Hood~II}
\affiliation{Kavli Institute for Cosmological Physics, University of Chicago, Chicago, IL, 60637, USA}
\affiliation{Department of Astronomy and Astrophysics, University of Chicago, Chicago, IL, 60637, USA}
\author{A.~Kosowsky}
\affiliation{Department of Physics and Astronomy, University of Pittsburgh, Pittsburgh, PA, 15260, USA}
\author{A.~La Posta}
\affiliation{Department of Physics, University of Oxford, Denys Wilkinson Building, Keble Road, Oxford OX1 3RH, United Kingdom}
\author{T.~Louis}
\affiliation{Universit\'e Paris-Saclay, CNRS/IN2P3, IJCLab, 91405 Orsay, France}
\author{M.~S.~Madhavacheril}
\affiliation{Department of Physics and Astronomy, University of Pennsylvania, 209 South 33rd Street, Philadelphia, PA, USA 19104}
\author{J.~McMahon}
\affiliation{Kavli Institute for Cosmological Physics, University of Chicago, Chicago, IL, 60637, USA}
\affiliation{Department of Astronomy and Astrophysics, University of Chicago, Chicago, IL, 60637, USA}
\affiliation{Fermi National Accelerator Laboratory, Batavia, IL, USA}
\affiliation{Enrico Fermi Institute, University of Chicago, Chicago, IL, USA}
\affiliation{Department of Physics, University of Chicago, Chicago, IL, USA}
\author{K.~Moodley}
\affiliation{Astrophysics Research Centre, University of KwaZulu-Natal, Westville Campus, Durban 4041, South Africa}
\affiliation{School of Mathematics, Statistics \& Computer Science, University of KwaZulu-Natal, Westville Campus, Durban 4041, South Africa}
\author{T.~Mroczkowski}
\affiliation{European Southern Observatory, Karl-Schwarzschild-Strasse 2, Garching 85748, Germany}
\affiliation{Institute of Space Sciences (ICE, CSIC), Carrer de Can Magrans, s/n, 08193 Cerdanyola del Vall\`es, Barcelona, Spain}
\author{S.~Naess}
\affiliation{Institute of Theoretical Astrophysics, University of Oslo, Norway}
\author{L.~Newburgh}
\affiliation{Department of Physics, Yale University, New Haven, CT 06520, USA}
\author{M.~D.~Niemack}
\affiliation{Department of Physics, Cornell University, Ithaca, NY 14853,USA}
\affiliation{Department of Astronomy,  Cornell University, Ithaca, NY 14850, USA}
\author{L.~A.~Page}
\affiliation{Joseph Henry Laboratories of Physics, Jadwin Hall, Princeton University, Princeton, NJ 08544, USA}
\author{B.~Partridge}
\affiliation{Department of Physics and Astronomy, Haverford College, Haverford, PA, 19041, USA}
\author{R.~Puddu}
\affiliation{Instituto de Astrof\'isica, Facultad de F\'isica, Pontificia Universidad Cat\'olica de Chile, Av. Vicu\~na Mackenna 4860, 7820436 Macul, Santiago, Chile}
\affiliation{Centro de Astro-Ingenier\'ia, Pontificia Universidad Cat\'olica de Chile, Av. Vicu\~na Mackenna 4860, 7820436 Macul, Santiago, Chile}
\author{E.~Schaan}
\affiliation{Kavli Institute for Particle Astrophysics and Cosmology, 382 Via Pueblo Mall Stanford, CA 94305-4060, USA}
\affiliation{SLAC National Accelerator Laboratory 2575 Sand Hill Road Menlo Park, California 94025, USA}
\author{N.~Sehgal}
\affiliation{Physics and Astronomy Department, Stony Brook University, Stony Brook, NY 11794, USA}
\author{C.~Sif\'on}
\affiliation{Instituto de F\'isica, Pontificia Universidad Cat\'olica de Valpara\'iso, Casilla 4059, Valpara\'iso, Chile}
\author{D.~N.~Spergel}
\affiliation{Simons Foundation, 160 5th Ave, New York, NY 10010, USA}
\author{S.~T.~Staggs}
\affiliation{Joseph Henry Laboratories of Physics, Jadwin Hall, Princeton University, Princeton, NJ 08544, USA}
\author{A.~van~Engelen}
\affiliation{School of Earth and Space Exploration,Arizona State University, Tempe, AZ 85287, USA}
\author{C.~Vargas}
\affiliation{Instituto de Astrof\'isica, Facultad de F\'isica, Pontificia Universidad Cat\'olica de Chile, Av. Vicu\~na Mackenna 4860, 7820436 Macul, Santiago, Chile}
\affiliation{Centro de Astro-Ingenier\'ia, Pontificia Universidad Cat\'olica de Chile, Av. Vicu\~na Mackenna 4860, 7820436 Macul, Santiago, Chile}
\author{E.~M.~Vavagiakis}
\affiliation{Department of Physics, Duke University, Durham, NC 27710, USA}
\affiliation{Department of Physics, Cornell University, Ithaca, NY 14853,USA}
\author{K.~Wagoner}
\affiliation{Department of Physics, North Carolina State University, Raleigh, NC 27695,USA}
\author{E.~J.~Wollack}
\affiliation{NASA Goddard Space Flight Center, Greenbelt, MD 20771, USA}

\date{\today}

\begin{abstract}
    \noindent
     The mean pairwise velocity of massive halos reflects the gravitational force law on cosmic scales. We combine cosmic microwave background intensity maps from the Atacama Cosmology Telescope and a galaxy catalog from the Sloan Digital Sky Survey to estimate the mean pairwise velocity using the kinematic Sunyaev-Zeldovich (kSZ) effect. On scales from 30 -- 230 megaparsecs, we constrain the gravitational acceleration between pairs of halos at separation $r$ to be $g\propto 1/r^n$ with $n=2.1\pm 0.3$, which is consistent with Newtonian gravity in an expanding spacetime (\textit{i.e.}, the standard $\Lambda$CDM model). This constraint shows agreement with an inverse quadratic radial dependence over the large distances separating galaxy halos, as expected in standard cosmology. Upcoming surveys have the potential to rule out $n = 1$ at $10\sigma$ significance. Our results establish the kSZ effect as a powerful tool for testing gravity on cosmological scales.
\end{abstract}

\maketitle

The standard cosmological model, $\Lambda$CDM, assumes that general relativity accurately describes gravitational interactions on all scales. This framework successfully explains a wide range of cosmic phenomena, from the galaxy correlation function to the cosmic microwave background (CMB). Usually, we can approximate the gravitational force law using Newtonian gravity, which falls off as the square of the separation between masses, $1/r^2$. However, the observed cosmic acceleration raises questions about whether this gravitational force law holds on the largest scales.
 
In addition to dark energy, several modified theories of gravity, including those that attempt to explain dark matter (DM), predict changes to the gravitational force law at large scales \citep{Clifton2012, Will2006}. 

The gravitational force law has been measured well within the solar system \citep[see, e.g.,][]{Adelberger2003}, but direct measures at larger scales are lacking. In addition, there are constraints on large scales \citep[see e.g., ][for some recent examples]{Chudaykin2025, Castello2022}; however, these all rely on an underlying, known homogeneous expansion history, which reduces the space of tested theories.

Galaxy and galaxy cluster velocities provide a promising avenue for testing the gravitational force law on cosmic scales. Specifically, pairwise velocities as a function of galaxy cluster separation provide a direct measurement of the acceleration field. Because the acceleration field is a function of both the matter field and the relationship between it and the resulting gravitational acceleration, pairwise velocities can test the gravitational force law in addition to growth of structure and dark energy.

The kinematic Sunyaev-Zeldovich effect \citep[kSZ;][]{Sunyaev1980} probes the baryonic velocity field at large scales. The pairwise kSZ effect was first measured using the Atacama Cosmology Telescope (ACT) maps combined with the Sloan Digital Sky Survey Data Release 9 (SDSS DR9) galaxy catalog \cite{Hand2012PhysRevLett.109.041101}, with many subsequent detections \cite{debernardis,calafut2021,Soergel2016, Schiappucci2023}. These recent measurements give us an opportunity to test gravity in a new way, at previously inaccessible scales.

In this \textit{Letter}, we use ACT measurements of the pairwise kSZ effect along with BOSS measurements of the correlation function to place constraints on the gravitational force law at cosmic scales. We discuss the implications for general modifications to the force law, as well as for one theory, Modified Newtonian Dynamics \cite[MOND;][]{Milgrom1983, Bekenstein1984} specifically (see the Appendix for a longer discussion of this theory). As we show, a modified gravitational force law that manages to explain the correlation function of galaxy clusters could still fail to predict the correct mean pairwise velocity. 

 \vspace{-0.5cm}
\section{Theory}\label{sec:theory}

\noindent In the following, we assume that the Universe is isotropic and homogeneous, and that it is expanding according to the Friedmann equations. Physical separations $r$ are related to comoving distances $x$ via the scale factor $a$, $r(t) = a(t) x(t)$. We do not specify a specific model of structure formation, but assume that any viable theory reproduces the measured galaxy correlation function at low redshifts. We also assume that halo peculiar velocities arise solely from gravitational accelerations with other halos. Throughout the following, we use `halos' to refer to galaxy groups or clusters. This formalism applies for galactic halos as well, but we mainly probe the velocities of galaxy clusters with the kSZ data used here.

\subsection{\texorpdfstring{$\Lambda$CDM}{LCDM} Derivation}
The pairwise velocity between any two halos in an expanding background is defined as
\begin{equation}\label{eqn:vpairdef}
    \mathbf{V}_{12} = \mathbf{v}_2 - \mathbf{v}_1 + \frac{\dot{a}}{a}\mathbf{r}_{12} \; ,
\end{equation}
where $\mathbf{v}_i$ is the peculiar velocity of each halo, and dots indicate proper time derivatives. Each halo itself follows the equation of motion
\begin{equation}\label{eqn:euler}
\frac{d\mathbf{v}_i}{dt} + \frac{\dot{a}}{a}\mathbf{v}_i= \mathbf{g}_i ,
\end{equation}
where $\mathbf{g}_i$ is the gravitational acceleration at the position of each halo.

The pairwise velocity can then be related to cosmological parameters. The typical derivation \citep{Peebles1976, Davis1977} uses the conservation of the number of halo pairs over short timescales to derive the mass-averaged pairwise velocity equation\footnote{We provide the full derivation in the apendix \cite{supplement_doiurl}.}
\begin{equation}\label{eqn:normal_pairwise}
    V(r, z) = -\frac{2}{3} H(z)rf(z) \frac{\bar{\xi}(r)}{1+\xi(r)} \; ,
\end{equation}
where $H(z) = \dot{a}/a$, $\xi(r)$ is the matter auto-correlation function, $\bar{\xi}(r) = \frac{3}{r^3}\int_0^r dr' r'^2\xi(r')$ is the spherically-averaged version, and $f(z)$ is the linear growth function.

\newcommand{\DOF}{15}
\newcommand{\LCDMCHISQ}{20.1}
\newcommand{\MONDCHISQ}{33.8}
\newcommand{\LCDMPTE}{0.17}
\newcommand{\MONDPTE}{3.65$\times 10^{-3}$}

\begin{figure*}[!t]
    \centering
    \includegraphics{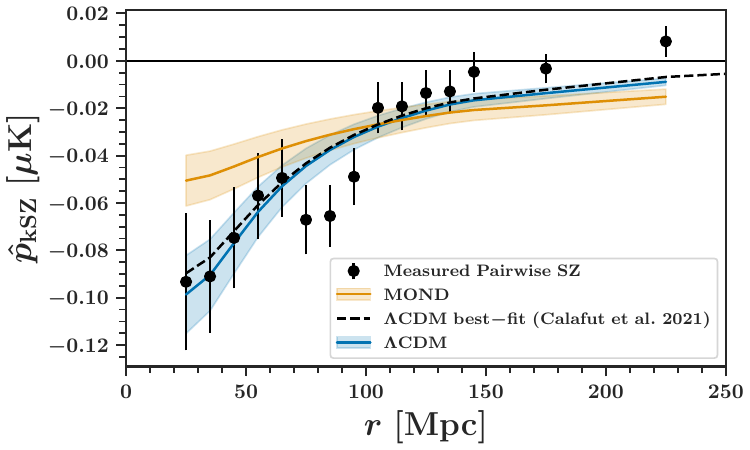}
    \caption{Pairwise kSZ measurements [$\mu$K] as a function of the physical separation of galaxy clusters [$\rm{Mpc}$]. Black dots with 1$\sigma$ errorbars show the measured pairwise momentum curve, obtained with the pairwise kSZ estimator and the ACT DR6 map. Blue line gives the amplitude-fitted $\Lambda \rm CDM$ prediction  using Eq.~(\ref{eqn:v12_fit}), with $n=2$, and the published SDSS correlation function. The orange line shows the best fit for MOND using Eq.~(\ref{eqn:v12_fit}), with $n=1$, for the same correlation function. Shaded regions indicate the 1$\sigma$ uncertainties on the amplitude. The dashed black line gives the $\Lambda \rm CDM$ prediction using a \Planck{} cosmology \citep{calafut2021}. Note that our $\Lambda$CDM fit closely matches previous work, which used a slightly expanded galaxy sample. MOND provides a poor fit to the data (PTE=\MONDPTE), while the $\Lambda$CDM case is in agreement (PTE=\LCDMPTE). The largest difference to the best-fit line is at 2.6$\sigma$, which is not statistically significant for the 15 bins shown.  A shift in the mean optical depth to these galaxy clusters would shift the amplitude of these fits, but not the shapes. }
    \label{fig:pksz}
\end{figure*}

However, this equation assumes scale-independent growth, which cannot be assumed generally. Here we instead express $V(r)$ in terms of local gravity, without an appeal to the growth of structure. Using Eqs.~(\ref{eqn:vpairdef}) \& (\ref{eqn:euler}), which do not rely on a specific gravitational theory, we can find the equation governing the pairwise velocity evolution. To first order in time derivatives, $\frac{d\mathbf{V}_{12}}{dt} = \mathbf{g}_2 - \mathbf{g}_1 \;$.

If we average over all pairs, we have the time derivative of the mean pairwise velocity, $\langle dV(r)/dt\rangle$, on the left hand side. The right hand side will be the mean pairwise acceleration, which we denote by $\langle g \rangle$ in the following. Note that these quantities are both scalars -- spherical symmetry at large scales enforces that both of these will depend solely on the magnitude of the pair separation.

The pairwise acceleration between a pair of halos at $\mathbf{r}_1$ and $\mathbf{r}_2$ is given by the difference in the gravitational accelerations at each point (\textit{i.e.}, the derivative of Eq.~(\ref{eqn:vpairdef})). To relate the pairwise acceleration to its sources, we consider the correlations of the pair with a third point, $\mathbf{r}_3$, which sources the gravitational force \citep{Peebles1980, Seto1999}. On average, this is
\begin{align}
        \hat{\delta}(\mathbf{r}_3;\mathbf{r}_1,\mathbf{r}_2) = &\frac{1}{1+\xi_{hh}(r_{12})} \Big[1+\xi_{hh}(r_{12}) \nonumber \\
        &+ \xi_{h}(r_{23}) + \xi_{h}(r_{31}) + \zeta_{hh}(r_{123}) \Big]\; ,
\end{align}
where $\xi_{hh}(r)$ is the halo auto-correlation function, $\xi_{h}(r)$ is the halo-matter cross-correlation function, and $\zeta_{hh}(r_{123})$ is the halo-halo-matter three point correlation function with sides given by the three considered points. In other words, this is the probability of finding matter at $\mathbf{r}_3$ and halos at $\mathbf{r}_{1,2}$, divided by the probability of finding halos at $\mathbf{r}_{1,2}$. We assume that this equation holds regardless of the gravitational force law. The mean pairwise acceleration in $\Lambda$CDM can then be formed from the gravitational acceleration due to the integrated mass from that density contrast, projected onto the line connecting the galaxy pair (since transverse accelerations average to zero) \citep{Seto1999},
\begin{equation}\label{eqn:full_gr}
   \langle g(r_{12}) \rangle = 2G\rho_m \int d\mathbf{r}_3\,\hat{\delta}(\mathbf{r}_3;\mathbf{r}_1,\mathbf{r}_2) \frac{\mathbf{r}_3 - \mathbf{r}_1}{|\mathbf{r}_3 - \mathbf{r}_1|^3} \cdot \mathbf{e}_{12} \; ,
\end{equation}
where $G$ is Newton's constant, $\rho_m$ is the average matter density in the Universe, $\mathbf{e}_{12} \equiv (\mathbf{r}_2 - \mathbf{r}_1) / r_{12}$, and $r_{12} = |\mathbf{r}_2 - \mathbf{r}_1|$. Note that the exact relation here assumes the Newtonian inverse square law relationship.

Combining results, we find that the mean pairwise acceleration in $\Lambda$CDM is\footnote{The simplification relies on symmetrizing the integral, and then solving the resulting integrals in Fourier-space, which includes a Bessel integral. We provide a full derivation in the Apendix.}
\begin{equation}\label{eqn:lcdm_v12}
    \left\langle \frac{dV}{dt} \right\rangle = \frac{4\pi G \rho_m}{r^2 (1+\xi_{hh}(r,t))} \int_0^{r} \xi_h(r',t) r'^2 dr'
\end{equation}
at leading order (e.g., ignoring higher-order correlations), and where we now make the correlation function time-dependence explicit. Note that if we assume scale-independent growth for the correlation functions, we can integrate this equation and exactly match Eq.~(\ref{eqn:normal_pairwise}).

\subsection{The Gravitational Force Law Generalization}
\noindent In the interest of full generality, we will take the modified gravity law
\begin{equation}
    g(r) = g_0\left[G_b(r)\right]^{n/2} \; ,
\end{equation}
where $g_0$ is a normalization factor with units of acceleration, $G_b(r)$ is a unitless measure of the baryonic acceleration (\textit{i.e.}, the acceleration due to baryonic matter alone divided by a normalizing acceleration to make it unitless), and $n$ is a power law index, with $n=2$ for $\Lambda$CDM. This parametrization for the force law index preserves the typical Newtonian proportionality, $g\propto 1/r^n$ with $n=2$ giving the Newtonian case. A slightly different case, in which the gravitational force power law is modified instead of the gravitational acceleration, is analyzed in the Appendix.

Our pairwise velocity evolution equation is then straightforward to compute given the same formalism as above. To make it easy to include modified gravity theories without dark matter, we will focus on the baryonic acceleration by using the galaxy auto-correlation function as a tracer of all correlation functions, though this leads to a small bias in both the  $\Lambda$CDM case and MOND. We find
\begin{equation}\label{eqn:mg_v12}
    \left\langle \frac{dV}{dt} \right\rangle = g_0 \left(\frac{4\pi G \rho_b/r^2}{\bar{a} (1+\xi_{gg}(r,t))} \int_0^{r} \xi_{gg}(r',t) r'^2 dr'\right)^{\frac{n}{2}} \; ,
\end{equation}
where $\rho_b$ is the baryonic matter density. We include $\bar{a} =  1\times 10^{-14}~\rm{m/s^2}$, as an arbitrary way to normalize the function. This is roughly the calculated pairwise acceleration in $\Lambda$CDM at the largest separations. This integral is dominated by scales $> 10~\rm{Mpc}$. Beginning our integration at different scales from $1-10~\rm{Mpc}$ introduces errors of $< 10\%$. Thus, any effects, such as galaxy bias, that introduce small shape changes at these scales should be absorbed by the multiplicative constant, $g_0$.

To solve the above differential equations for the pairwise velocity, we need the time-dependence of the correlation function. We circumvent this by considering a narrow slice in time (\textit{i.e.}, galaxy redshift) over which to solve for the velocity. As we discuss below, the correlation function only changes $\sim 2\%$ during this redshift slice, and so taking a constant $\xi_{gg}(r,t)$ over this redshift slice should be a good approximation.

Using this simplification, and substituting in the spherically averaged correlation function, $\bar{\xi}_{gg}(r)$, we find
\begin{equation}\label{eqn:v12_fit}
    V(r) = A \left(\frac{4\pi G \rho_b r/\bar{a}}{1+\xi_{gg}(r)}\bar{\xi}_{gg}(r)\right)^{n/2} \; ,
\end{equation}
where $A = g_0 \Delta t$, and $\Delta t$ is the arbitrary, small time interval over which the equation is solved. The $n=2$, $A=\bar{a}\Delta t\rho_m/\rho_b$ case corresponds to $\Lambda$CDM, and agrees with our Eq.~(\ref{eqn:lcdm_v12}), up to bias factors in the correlation function. Note that we will not use the amplitude, $A$, in our full analysis. We instead replace it with the mass-averaged optical depth to the galaxy sample, $\bar{\tau}$. As shown below, when we consider our actual observable, $A$ is degenerate with $\bar{\tau}$.

In the appendix, we derive how MOND fits into this formalism. Our general equation, Eq.~(\ref{eqn:v12_fit}) with $n=1$, is appropriate for MOND in the large-scale limit.

\section{Observables \& Analysis}

\subsection{Pairwise Velocity Inference from the Galaxy Correlation Function}
\noindent Using the SDSS Baryon Oscillation Spectroscopic Survey galaxies as tracers of halos, we use the galaxy correlation function $\xi_{gg}$, to evaluate the right-hand side of Eq.~(\ref{eqn:v12_fit}).\cite{ross_2016, sdss_dataproducts, alam_10.1093/mnras/stx721}  To constrain the time-evolution of the correlation function, we use the redshift bin in the range $0.44<z<0.66$ according to Refs. \cite{ross_2016, alam_10.1093/mnras/stx721}, which includes the mean redshift of the full catalog. This redshift bin consists of 686k galaxies. We expect the growth function to change $\sim 15\%$ during this redshift interval, and thus the galaxy correlation function only changes by $\sim 2\%$. We use the mean values and covariances of the two-point correlation function to simulate 1,000 random realizations of $\xi_{gg}$. We carry out the integral to calculate the spherically-averaged correlation function in Eq.~(\ref{eqn:v12_fit}) using a lower limit of $6 \, \rm{Mpc}$, which gives a difference of $3-7\%$ between 30 -- 100 Mpc compared to the full integral, assuming linear theory. We compute Eq.~(\ref{eqn:v12_fit}) and its covariance using this prescription. We use the \Planck{} 2018 cosmology \cite{Planck2018} to set $\rho_b= 4.21\times 10^{-31}~\rm{g/cm}^3$. Note that any value could be used here, since it would be absorbed by the overall normalization factor.

\subsection{Pairwise kSZ}
\label{subsec:pairwiseksz}
\noindent CMB experiments measure the pairwise momentum, which is related to the mass-averaged pairwise velocity by \begin{equation}\label{eqn:pksz}
    \hat{p}(r, z) = -\frac{T_{ \rm CMB}}{c} \bar \tau V(r, z) \; ,
\end{equation}
where $\bar \tau$ is the effective mass-averaged optical depth for the group/cluster sample, and $z$ is the mean redshift. As a reminder, we replace the amplitude of the pairwise velocity with $\bar \tau$ in what follows.

We use the pairwise kSZ estimator as a measure of the pairwise velocities of groups/cluster as traced by a red luminous sample of galaxies \cite{Ferreira1999}, following the methods in Ref.~\cite{calafut2021}. The pairwise estimator,
\begin{equation}
    \label{eq:pairwise}
    \hat p_{\rm kSZ}(r) = -\frac{\sum_{j>i} c_{ij} (dT_i-dT_j)}{\sum_{j>i} c_{ij}^2} \; ,
\end{equation}
consists of the pair differences of redshift-corrected temperature decrements obtained with an aperture photometry filter (with $r_{disk}=2.1 \, \rm{arcmin}$ and $r_{ring}=\sqrt{2}r_{disk}$, as in Ref. \cite{calafut2021}) centered on a galaxy tracer. The line of sight weights, $c_{ij}=\hat {\textbf{r}}_{ij} \cdot \frac{\hat{ {\textbf{r}}}_i + \hat{ \textbf{r}}_j}{2}$, are determined using on-sky positions and redshifts from a spectroscopic galaxy catalog. The sum in Eq.~(\ref{eq:pairwise}) is carried out for all pairs of tracer galaxies from a catalog that lie within a separation bin \cite{Gallardo_2025_iskay}. 

We use the ACT coadd map from the most recent Data Release 6, which combines ACT 150 GHz with \Planck{} 143 GHz temperature maps (\cite{Naess_2025, Naess_2020}). This map passed numerous data splits designed to limit systematic errors \cite{Naess_2025, louis2025atacamacosmologytelescopedr6}. In addition, single frequency maps have been shown to be robust against selection effects as shown in Ref.~\cite{calafut2021}. Red luminous galaxies with spectroscopic redshifts from the SDSS DR 15 \cite{Aguado_2019} are used with k-corrected luminosities $L>4.3 \times 10^{10}L_{\odot}$, and we mask according to \cite{vavagiakis21}. In contrast to Ref.~\cite{calafut2021}, we further restrict the sample to a small redshift slice, $0.44<z<0.66$, matching the SDSS sample we use for the correlation function. The total sample contains 343,647 galaxies, where 227,837 galaxies are in the selected redshift range. The pairwise kSZ covariance matrix and photometry aperture are computed according to Ref.~\cite{calafut2021}.

\newcommand{\BESTFITN}{2.1}
\newcommand{\SIGMAN}{0.3}
\newcommand{\NSIGMAMOND}{3.3}
\newcommand{\NSIGMALCDM}{0.4}
\newcommand{\NLOWERBOUNDNINETYFIVEPCT}{1.4}

\begin{figure}
    \centering    \includegraphics[width=\columnwidth]{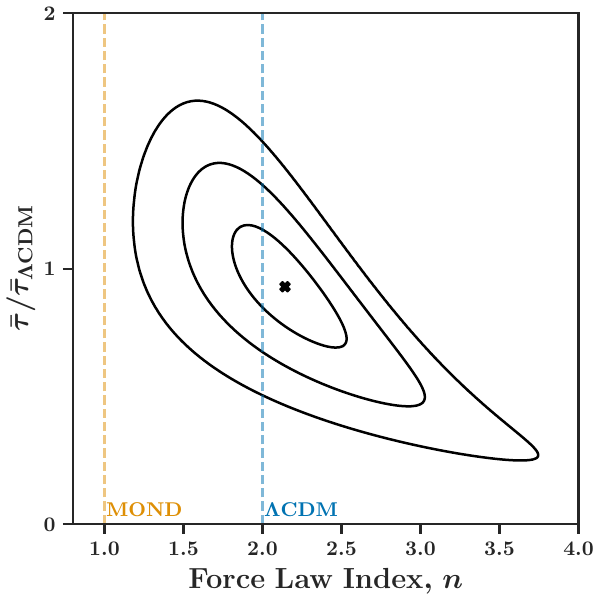}
    \caption{Constraints on the gravitational force law index, $n$, and the average optical depth $\bar \tau$ from the ACT DR6 $150\, \rm GHz$ map pairwise kSZ measurement. Black lines give the 1,2,3$\sigma$ bounds. Wide and uniform priors were assigned to $n$ and $\bar\tau$. The average optical depth ratio is obtained using in Eq.~(\ref{eqn:v12_fit}), the CMB temperature, $T_{\rm{CMB}}$, and the inferred standard cosmology optical depth ($\bar \tau_{\Lambda CDM}$) as in C21 Table II. The marginalized value for the force law index is $n=\BESTFITN\pm \SIGMAN$. We include dashed lines to indicate the $\Lambda$CDM (blue) and MOND (orange) force law parameters. The $y$-axis scale of the plot is dependent on the choice of the arbitrary normalization $\bar{a}$ in Eq.~(\ref{eqn:v12_fit}); however, the extent across $n$ is not and the maximum likelihood position remains fixed regardless of $\bar{a}$. MOND is outside of our 3$\sigma$ contour.}
    \label{fig:contour}
\end{figure}

\subsection{Analysis}
\label{subsec:analysis}

\noindent We use two tests to probe the agreement between theory and observations. First, we fix the force law index $n$ and fit an amplitude to the theoretical prediction for the velocity curve. In concordance cosmology, this amplitude term is proportional to the average optical depth $\bar \tau$. Second (and more generally), we jointly fit the amplitude and force law index by extremizing the likelihood function.

To evaluate the theoretical predictions described above, we compute the quantity 
$\chi ^2 = \sum_{ij} \Delta \hat p_{i} \Sigma^{-1}_{ij} \Delta \hat p_{j}$, where $\Delta \hat p_{i} = \hat p_{i, \rm data} - \hat p_{i, \rm theory}$, and $\Sigma_{ij}=\Sigma_{ij}^{\rm{kSZ}} + \Sigma_{ij}^{\rm{SDSS}}$ is the combined covariance matrix using the measured bootstrapped pairwise kSZ covariance and the covariance from random realizations of the SDSS correlation function. 
Eq.~(\ref{eqn:v12_fit}) is used to fit a linear amplitude to the radial dependence of the measured pairwise curves in the case of $\Lambda$CDM ($n=2$) and a MOND force law ($n=1$). We evaluate the probability to exceed, $\mathrm{PTE} = \int_{\chi^2_{\rm min}} ^\infty p_m(x) dx$,
 where $p_m(x)$ is the probability density function for a $\chi^2_m$ distribution with $m$ degrees of freedom. 

In the more general case, we evaluate the likelihood $-2 \mathrm{log}(\mathcal{L}) = \sum_{ij} \Delta \hat p_i \Sigma^{-1}_{ij} \Delta \hat p_j$, where the theoretical prediction follows a two variable model. Wide and uniform priors are assigned to  the amplitude (parameterized by the optical depth, $\bar{\tau}$) and force law index ($n$). In this case, the covariance matrix $\Sigma_{ij}$ is approximated by the covariance of the pairwise measurement, which dominates the uncertainty, to simplify the calculation.

\section{Results \& Discussion}

\noindent Following the above prescription, we evaluate the fit of different gravitational force laws to the data for both fixed and free force law index, $n$. The best-fit values for the force law index $n=2$ ($\Lambda$CDM) and $n=1$ (MOND) give $\chi^2_{\rm min}/N_{\rm DoF} =$ \LCDMCHISQ/\DOF~(PTE=\LCDMPTE) and \MONDCHISQ/\DOF~(PTE=\MONDPTE) respectively, indicating that MOND fits the pairwise velocities poorly. The Newtonian gravity case fits the data well, agreeing with the $\rm \Lambda CDM$ model. The best-fit curves for each model are shown in Fig.~\ref{fig:pksz}. The $\Lambda \rm CDM$ velocity curve estimated using our formalism is shown in blue, and the MOND best fit case is shown in orange. The color bands show the $1\sigma$ uncertainty on the fitted amplitude. The black, segmented line shows the $\rm \Lambda CDM$ theoretical prediction using Eq.~(\ref{eqn:normal_pairwise}) with \Planck{} cosmological parameters \cite{planck_parameters_theory_curve}, as fitted in Ref.~\cite{calafut2021} for the full galaxy sample with $L>4.3\times 10^{10} L_\odot$. Note that our $\Lambda$CDM prediction fits that of previous work well \cite{calafut2021}.

The posterior for the joint two-parameter fit (constraining the optical depth $\bar{\tau}$ and force law index $n$) is shown in Fig.~\ref{fig:contour}, where contours show $1,2,3\sigma$ intervals. We normalize $\bar \tau$ by the best-fit $\Lambda$CDM as in Ref.~\cite{calafut2021}, excluding corrections as in Ref. ~\cite{Gong:2023hse}. The marginalized estimate of the power law index is $n = \BESTFITN \pm \SIGMAN$ ($68\%$ C.I.), which deviates from the MOND force law index ($n=1.0$) at $\NSIGMAMOND \sigma$, and from the $\Lambda$CDM (n=2) value at $\NSIGMALCDM \sigma$. This constrains the force law index to $n>\NLOWERBOUNDNINETYFIVEPCT$ at 95\% C.I.

Our analysis here is quite general -- it does not rely on a specific cosmology. However, we assume that there are no large changes to the correlation function over the small redshift range we are probing, which agrees with the $\Lambda$CDM prediction. If the observed correlation function today is to remain the same, then breaking the assumption of a slowly changing correlation function at $z=0.44-0.66$ would induce large changes in the observed correlation function at other redshifts, in addition to likely deviations from the observed growth rate \citep[e.g.,][]{Alam2021}.

Although the bulk of our analysis is model-independent, we have also shown the theoretical curves appropriate for MOND. This test is the largest-scale direct test of MOND to date. Our formalism is an excellent approximation to MOND in the low-acceleration regime; that said, we have not included the external field effect (EFE) in this analysis \citep{McGaugh2013}. This can modify the acceleration law in the case where the baryonic gravity of an object is less than the gravity of its larger environment, and has been used to explain the velocity dispersion of satellite galaxies \citep{McGaugh2013, Safarzadeh2021}. However, this effect likely does not affect our analysis. The smallest separation we consider is $\sim 30~\rm{Mpc}$, which can be compared to the largest virialized objects in the universe (galaxy clusters), with maximum sizes of $\sim 10~\rm{Mpc}$. There is one caveat to this statement: our analysis includes an integration over the correlation function to scales smaller than $30~\rm{Mpc}$. As noted below Eq.~(\ref{eqn:mg_v12}), this integration is still largely dominated by scales larger than $ 10~\rm{Mpc}$. A full exploration of whether the EFE could affect the results here would require a simulation of the EFE over the scales we probe.

\newcommand{\CURRENTNOISE}{14}
\newcommand{\CURRENTNGAL}{227,837}
\newcommand{\NOISEFORECAST}{6}
\newcommand{\NGALFORECASTMILLIONS}{4}
\newcommand{\SNRFORECASTEDSO}{10}

High precision maps from future CMB experiments combined with ongoing optical spectroscopic surveys (which will provide large galaxy catalogs) will provide high-resolution kSZ measurements, thus enabling stringent tests of alternative models of gravity. In particular, using a simple scaling $\sigma\propto{1}/{\sqrt{N_{\rm gal}}}$, as in Ref. \cite{ried_etal_lqbj-wcqj},  using \NGALFORECASTMILLIONS{} million galaxies \cite{SO_forecasts, DESICollaboration2016} and the current statistical significance of $3.3\, \sigma$, we forecast that the test presented here could rule out $n=1$ at $\SNRFORECASTEDSO \sigma$. Simulations suggest that this scaling is adequate for DESI but will overestimate the SNR for LSST due to cosmic variance \cite{schutt_PhysRevD.109.103539}.

\vskip 4pt
In summary, this paper uses the kinematic Sunyaev-Zeldovich (kSZ) effect to place constraints on the form of the gravitational force law at cosmological scales. The kSZ effect probes the pairwise velocities of galaxies, which can be related to a gravitational force law via the galaxy correlation function. This paper develops this formalism in a model-independent way, and uses ACT DR6 to perform the test. We show that $\Lambda$CDM fits the data well, and that any deviations to the gravitational force law must be constrained to power law indices of $n = \BESTFITN \pm \SIGMAN$ ($68\%$ C.I.; marginalized over $\bar{\tau}$). In addition, we showed that MOND specifically is excluded at $>3\sigma$. With the currently observing Simons Observatory and its upgrades \cite{SO_forecasts, ASO_Abitbol_2025}, the kSZ data will improve significantly, and our approach can be used to test the gravitational force law to even greater fidelity.

\acknowledgments
\section{Acknowledgments}
The authors would like to thank the anonymous referees for their useful feedback. This manuscript also benefited from discussions with Harry Desmond, Pedro Ferreira, Daniel Holz, and Tariq Yasin. PAG acknowledges support from the Kavli Institute for Cosmological Physics at the University of Chicago. OHEP is a Junior Fellow of the Simons Society of Fellows. RB acknowledges support from NSF grant AST-2206088 and NASA ROSES grant 12-EUCLID12-0004.  RD, CH, CS, CV acknowledge support from the Agencia Nacional de Investigaci\'on y Desarrollo (ANID) through Basal project FB210003. KM and MH acknowledge support from the National Research Foundation of South Africa. ADH acknowledges support from the Sutton Family Chair in Science, Christianity and Cultures, from the Faculty of Arts and
Science, University of Toronto, and from the Natural Sciences and Engineering Research Council of Canada (NSERC) [RGPIN-2023-05014, DGECR-2023-00180]. Support for ACT was through the U.S.~National Science Foundation through awards AST-0408698, AST-0965625, and AST-1440226 for the ACT project, as well as awards PHY-0355328, PHY-0855887 and PHY-1214379. Funding was also provided by Princeton University, the University of Pennsylvania, and a Canada Foundation for Innovation (CFI) award to UBC. ACT operated in the Parque Astron\'omico Atacama in northern Chile under the auspices of the Agencia Nacional de Investigaci\'on y Desarrollo (ANID). The development of multichroic detectors and lenses was supported by NASA grants NNX13AE56G and NNX14AB58G. Detector research at NIST was supported by the NIST Innovations in Measurement Science program. Computing for ACT was performed using the Princeton Research Computing resources at Princeton University, the National Energy Research Scientific Computing Center (NERSC), and the Niagara supercomputer at the SciNet HPC Consortium. SciNet is funded by the CFI under the auspices of Compute Canada, the Government of Ontario, the Ontario Research Fund-Research Excellence, and the University of Toronto. We thank the Republic of Chile for hosting ACT in the northern Atacama, and the local indigenous Licanantay communities whom we follow in observing and learning from the night sky.

\section{Code Availability \& Software Packages}

The code used to produce our analysis and plots is publicly available at: \url{https://github.com/patogallardo/pairwiseksz_mond}. The following software packages were used throughout this study: \textit{Matplotlib} \cite{matplotlib}, \textit{Numpy} \cite{numpy}, \textit{Pandas} \cite{pandas, mckinney-proc-scipy-2010}, \textit{Scipy} \cite{scipy}, \textit{Iskay2} \cite{Gallardo_2025_iskay}.

\section{Appendix}
\appendix

\setcounter{equation}{11}

\section{Pairwise Velocity from Conservation of Pairs}\label{app:lcdm}
Here we briefly summarize how the pairwise velocity is typically derived in $\Lambda$CDM. We follow Refs. \cite{Peebles1976, Davis1977} in the following discussion.

We start by considering the definition of the pairwise velocity (Eq.~(\ref{eqn:vpairdef})) and the equations of motion for the galaxies (Eq.~(\ref{eqn:euler})), and then impose spherical symmetry. This constrains the two-point correlation function to be a function of only the magnitude of the pair separations. The average number of pairs of galaxies separated by at most $r$ is
\begin{equation}
    N(r,t) = N_0\int_0^{r}dr'~2\pi r'^2 [1+\xi(r', t)] \; ,
\end{equation} 
where $N_0$ is a normalization factor, and $\xi(r,t)$ is the matter auto-correlation function. Within a given small time interval, $\delta t$, the total number of pairs should not change (ignoring mergers, which happen on much longer timescales), giving the continuity equation
\begin{equation}
    \frac{\partial}{\partial t} \int_0^r dr' r'^2 \xi(r',t) = -r^2[1 + \xi(r, t)]V/a \; ,
\end{equation}
where $V \equiv \|\mathbf{V}\|$ is the mass-averaged pairwise velocity.

In terms of the spherically averaged correlation function, $\bar{\xi}(r) = \frac{3}{r^3}\int_0^r dr' r'^2\xi(r')$, this is:
\begin{equation}
    \frac{\partial \bar{\xi}}{\partial t} = -3\frac{V}{ar} [1 + \xi(r, t)] \; .
\end{equation}

The solution for the pairwise velocity depends on the time evolution of the correlation function, which is dominated by large scales. At late times, it is assumed that this can be described by a growth function, which describes a scale-free increase in structure \citep[cf.][]{Peebles1976, Juszkiewicz1999, Sheth2001}. This leads to the typical pairwise velocity formula \citep[as seen in, e.g.,][]{Juszkiewicz1999, Sheth2001}
\begin{equation}
    V = -\frac{2}{3} Hrf(z) \frac{\bar{\xi}(r)}{1+\xi(r)} \; ,
\end{equation}
where $f(z)$ is the linear growth function.

\section{Derivation of the Pairwise Velocity Equation}\label{app:vel_eqn_deriv}

Here we provide the full derivation for the main pairwise equations, Eqns.~\ref{eqn:full_gr} \& ~\ref{eqn:lcdm_v12}. 

First, we compute the peculiar acceleration experienced by an observer at comoving position $\vec x$ due to a distribution of matter with overdensity $\delta(\vec y)$. Assuming Newtonian gravity, $\vec g(\vec x, t) = -\nabla \phi(\vec x,t)/a(t)$, where the peculiar potential satisfies
\begin{equation}
    \nabla^2\phi(\vec x, t) = \frac{3}{2}H^2(t)a^2(t)\Omega_m(t)\delta(\vec x, t),
\end{equation}
with the standard (linear-order) solution
\begin{equation}
    \vec g(\vec x, t) = \frac{3}{8\pi}H^2(t)a(t)\Omega_m(t)\int d\vec y\,\delta(\vec y, t)\frac{\vec y-\vec x}{|\vec y-\vec x|^3}.
\end{equation}
Consider now two points at $\vec x_{1,2}$. Their pairwise peculiar acceleration is given by
\begin{eqnarray}
    g(\vec x_1,\vec x_2) &\equiv& \left[\vec g(\vec x_1, t)-\vec g(\vec x_2, t)\right]\cdot\vec e_{12} \nonumber \\\nonumber
    &=& \frac{3}{8\pi}H^2(t)a(t)\Omega_m(t)\int \Bigg( d\vec x_3\,\delta(\vec x_3, t) \\
    &&\times \left[\frac{\vec x_3-\vec x_1}{x_{13}^3}-\frac{\vec x_3-\vec x_2}{x_{23}^3}\right]\cdot\vec e_{12} \Bigg),
\end{eqnarray}
where $x_{ij}\equiv |\vec x_j-\vec x_i|$ and we have projected along the line-of-sight of the galaxy pair $\vec e_{12}\equiv (\vec x_1-\vec x_2)/x_{12}$ (note that any transverse motions average to zero). This is similar to Ref.~\citep{Seto1999}, but goes beyond the Einstein de-Sitter approximation. 

To obtain the mean pairwise acceleration, we must average over the distribution of matter, conditioning on the fact that there are galaxies at $\vec x_1$ and $\vec x_2$:
\begin{equation}
    \delta(\vec x_3) = \frac{1+\xi_{gg}(x_{12})+\xi_{g}(x_{13})+\xi_g(x_{23})+\zeta_{gg}(x_1,x_2,x_3)}{1+\xi_{gg}(x_{12})}
\end{equation}
following \cite{Peebles1980}, and leaving the time dependence implicit from here. This is simply the (rescaled) probability of finding \textit{matter} at $\vec x_3$ and galaxies at $\vec x_{1,2}$, divided by the probability of finding \textit{galaxies} at $\vec x_{1,2}$. In practice, these are averaged over a mass range, which modifies the linear bias factors in the cross- and auto- correlation functions $\xi_{g}$ and $\xi_{gg}$.

Our next step is to simplify the expression. At linear order, the three-point function $\zeta$ vanishes; furthermore, the first term (unity) in $\delta(\vec x_3)$ vanishes upon integration over $\hat{\vec x}_3$. This leaves
\begin{eqnarray}
    g(\vec x_1,\vec x_2) &=& \frac{3}{4\pi}H^2a\,\Omega_m \int \Bigg( d\vec x_3\,\frac{\xi_g(x_{13})+\xi_g(x_{23})}{1+\xi_{gg}(x_{12})} \nonumber \\
    &&\times \frac{\vec x_3-\vec x_1}{x_{13}^3}\cdot\vec e_{12} \Bigg),
\end{eqnarray}
where we note that the two terms ($\vec g(\vec x_1)$ and $\vec g(\vec x_2)$) are equal (seen under the redefinition $\vec x_3\to \vec x_3'\equiv\vec x_3+\vec x_2-\vec x_1$). Next, we note that the $\xi_g(x_{13})$ term in the above expression is zero: upon redefining $\vec x_3\to \vec x_3'=\vec x_3-\vec x_1$, the term is proportional to $\int d\hat{\vec{x}}_3'~\hat{\vec{x}}_3' = 0$. Noting that $g$ can depend only on the relative displacement $\vec x_1-\vec x_2\equiv \vec x$ by homogeneity and making the $\vec x_3\to \vec y=\vec x_3-\vec x_1$ substitution again, we find
\begin{eqnarray}
    g(\vec x) = \frac{3}{4\pi}H^2a\,\Omega_m\int d\vec y\,\frac{\xi_g(|\vec x-\vec y|)}{1+\xi_{gg}(x)}\frac{\vec y}{y^3}\cdot\hat{\vec x}.
\end{eqnarray}
Rewriting the correlation function in Fourier-space gives
\begin{eqnarray}
    g(\vec x) = &&\frac{3}{4\pi}\frac{H^2a\,\Omega_m}{1+\xi_{gg}(x)}\int d\vec y\, \nonumber \\
    &&\times \int\frac{d\vec k}{(2\pi)^3}P_g(k)e^{i\vec k\cdot(\vec x-\vec y)}\frac{\vec y}{y^3}\cdot\hat{\vec x}.
\end{eqnarray}
To simplify further, we note that
\begin{eqnarray}
    \int\frac{d\vec k}{(2\pi)^3}P_g(k)e^{i\vec k\cdot(\vec x-\vec y)} = &&\sum_\ell (2\ell+1)L_\ell(\hat{\vec x}\cdot\hat{\vec y}) \int\frac{k^2dk}{2\pi^2} \nonumber \\
    &&\times P_g(k)j_\ell(kx)j_\ell(ky),
\end{eqnarray}
for Legendre polynomial $L_\ell$ and spherical Bessel function $j_\ell$. Next, we can perform the $\vec y$ integral:
\begin{eqnarray}
    \int_0^\infty dy\,j_\ell(ky)&\int& d\hat{\vec y}\,L_\ell(\hat{\vec x}\cdot\hat{\vec y})\,\hat{\vec x}\cdot\hat{\vec y} \nonumber \\
    &=& \frac{4\pi}{3}\delta^{\rm K}_{\ell 1} \int_0^\infty dy\,j_1(ky) \nonumber \\
    &=& \frac{4\pi}{3k}\delta^{\rm K}_{\ell 1}.
\end{eqnarray}
This yields
\begin{eqnarray}
    g(x) = 3\frac{H^2a\,\Omega_m}{1+\xi_{gg}(x)}\int_0^{\infty}\frac{k^2dk}{2\pi^2}\frac{P_g(k)}{k}j_1(kx).
\end{eqnarray}
Finally, we can rewrite $P_g(k)$ in configuration-space as $P_g(k) = 4\pi \int_0^\infty y^2dy\,\xi_g(y)j_0(ky)$; noting that
\begin{eqnarray}
    \int_0^\infty\frac{k^2dk}{2\pi^2}\frac{j_1(kx)j_0(ky)}{k} =  \frac{1}{4\pi x^2} \quad \text{if } x\geq y
\end{eqnarray}
and zero else. This follows from a standard integral for cylindrical Bessel functions. Collecting results, we find
\begin{eqnarray}
    g(x) = \frac{H^2a\,\Omega_m}{1+\xi_{gg}(x)}\frac{3}{x^2}\int_0^x y^2 dy\,\xi_g(y).
\end{eqnarray}
Converting from comoving to physical coordinates and converting from Hubble's factor and $\Omega_m$ to matter density, this then matches Eqn.~\ref{eqn:full_gr} in the main text.

As described in the main text, we modify $g(r) = g_0(G_b(r))^{n/2}$ for the generalized gravity case. The above derivation is unchanged for this case, except for an overall modification of raising all equations to a power of $n/2$. 

\section{An Example: MOND}\label{app:MOND}
Modified Newtonian Dynamics (MOND) \citep{Milgrom1983, Bekenstein1984} postulates that the gravitational force law changes at very weak acceleration scales. This class of theories has been successful in predicting galaxy rotation curves for larger and Milky Way-sized galaxies \citep[see Refs.~][for reviews]{Sanders2002, Kosowsky2010, Famaey2012}, and the relationship between the observed velocity curves of galaxies and their baryonic content \citep[i.e., the radial acceleration relationship, ][]{McGaugh2004, McGaugh2005, McGaugh2012}. However, these modified gravity theories are highly constrained. MOND struggles to explain measurements of gravitational wave propagation \citep{Bekenstein2004, Abbott2017, LIGOGRTests2019, Boran2018}, to reproduce the diversity of dwarf galaxy velocity dispersions \citep{Nipoti2008, McGaugh2010, Safarzadeh2021}, to explain the gravitational mass within galaxy clusters \citep{Sanders1999, Sanders2003, Tian2020}, and to explain the growth of structure from the CMB to today \cite{Dodelson2011, Pardo2020}. Nevertheless, some of these constraints could be avoided via additional theoretical parameters \citep[e.g.,][]{skordis_PhysRevLett.127.161302}. In addition, all of these constraints are either indirect measures of gravity or do not measure gravity at scales where perturbations of the potential are small. In this paper we present a direct test via pairwise velocities. In the rest of this Appendix, we provide a derivation of the pairwise velocity equation for MOND and show that it maps onto the parametrization we use in the main text (Eq.~(\ref{eqn:v12_fit})).

In the case of MOND, the Poisson equation is modified \citep{Bekenstein2004} to $\nabla \cdot [\mu(|\textbf{g}_{\rm{M}}|/a_0) \mathbf{g}_{\rm{M}}] = 4\pi G\rho_b$, 
where $\mu(x)$ is a fitting function, $a_0  = 10^{-10}~\rm{m/s}^2$ \citep{Milgrom1983, Famaey2012} is a special acceleration scale, $g_{\rm{M}}$ is the gravitational acceleration predicted by MOND, and $\rho_b$ is the baryon density. The fitting function can have several forms; however, the limiting behavior is defined as: $\mu(x) \rightarrow 1$ for $x \gg 1$ and $\mu(x) \rightarrow x$ for $x \ll 1$. This is a nonlinear equation that usually requires numerical solution, and can lead to complex dynamics (e.g., the external field effect \citep{McGaugh2013}). 

We can instead make progress by following Bekenstein \& Magueijo and formally linearizing the equation \citep[see,][and e.g. \citealp{Milgrom1986}]{ Bekenstein2006}. In particular, we define the vector $\mathbf{u} = \mu \left(\frac{|\mathbf{g}_{\rm{M}}|}{a_0}\right) \mathbf{g}_{\rm{M}}$ that transforms the modified Poisson equation into two first-order equations
\begin{eqnarray}
    \nabla \cdot \mathbf{u} = 4\pi G\rho \label{eqn:fo_mond}\\
    \nabla \times \left(\frac{\mathbf{u}}{\mu}\right) = 0 \; . \label{eqn:curl}
\end{eqnarray}
Using this set of equations, $\mathbf{u}$ can be solved for each source in a given mass distribution, summed to find a total $\mathbf{u}$, and then inverted (via its definition) to get the MOND acceleration $\mathbf{g}_{\rm{M}}$.

Notably, Eq.~(\ref{eqn:fo_mond}) is equivalent to the Newtonian Poisson equation but with $\mathbf{u}$ taking the place of the Newtonian acceleration. For problems with planar symmetry, such as the above derivation (since we only consider three mass sources at one time), $\mathbf{u}$ is curl-free \citep{Bekenstein2006}. This allows Eq.~(\ref{eqn:fo_mond}) to be solved via the same manner as for Newtonian gravity and then related to the MOND potential via our definition of $\mathbf{u}$. We note that for more general theories, we should still expect the potential to be curl-free given the strong constraints on non-Gaussianity in large-scale structure \citep{Slosar2008, Cabass:2022wjy, Cabass:2022ymb}.

In this paper we only consider large scales ($\gtrsim 30~\rm{Mpc}$),\footnote{We have checked explicitly that is the case for our regime -- at most, the gravitational acceleration is $\sim 10^{-3} a_0$.} where $|\mathbf{g}_{\rm{M}}| \ll a_0$. In this ``deep MOND'' regime, $\mu \approx |g_{\rm{M}} | / a_0$, which gives $\mathbf{u} = \frac{|g_{\rm{M}}|}{a_0} \mathbf{g}_{\rm{M}}$. Setting the presumed Newtonian acceleration due to baryons $\mathbf{g}_b = \mathbf{u}$, and specializing to the deep MOND regime,
\begin{equation}\label{eqn:gmond}
     \mathbf{g}_{\rm{M}} = \sqrt{a_0 g_b} \hat{\mathbf{g}}_b.
\end{equation}

Next, we must find the average pairwise $\langle g_{\rm{M}}\rangle$. From Eq.~(\ref{eqn:gmond}), we see that $\langle g_{\rm{M}}\rangle = \langle \sqrt{a_0 g_b}\rangle$. We can Taylor expand this as $\langle g_{\rm{M}}\rangle = \sqrt{a_0\langle g_b(r)\rangle} + \mathcal{O}(\zeta(\mathbf{r}_i))$.
As above, we will ignore the three-point correlation function. We note that even if MOND had a much larger three-point function than $\Lambda$CDM at these large scales, it would be strongly constrained by large-scale structure measurements of non-Gaussianity \citep{Slosar2008, Cabass:2022wjy, Cabass:2022ymb}.

This gives a straightforward modification to Eq.~(\ref{eqn:full_gr}). The MOND equation is now
\begin{equation}
   \langle g_{\rm{M}}(r) \rangle = \left(2G\rho_ba_0 \int d\mathbf{r}_3 \hat{\delta}(\mathbf{r}_3;\mathbf{r}_1,\mathbf{r}_2) \frac{\mathbf{r}_3 - \mathbf{r}_1}{|\mathbf{r}_3 - \mathbf{r}_1|^3} \cdot \mathbf{e}_{12}\right)^{1/2}  \; ,
\end{equation}
where we replace $\rho_m$ with $\rho_b$, given that we only have baryons in the case of MOND. The integral is then simplified in the same way as in the $\Lambda$CDM case, and we are left with Eq.~(\ref{eqn:mg_v12}), where $n=1$. 

\section{An Alternative Model}
In main text, we use a model where the $n=1$ case corresponds to MOND. While this model is easier to compare to MOND, it does also imply that, for $n\neq 2$, gravity does not depend on matter density linearly. Here we consider an alternate model where we keep the linear matter density dependence and only consider a varying radial dependence 
    \begin{equation}
        \label{eq:altgraveq}
        \langle g(r_{12}) \rangle = 2G\rho_m R_0^{n-2} \int d\mathbf{r}_3\,\hat{\delta}(\mathbf{r}_3;\mathbf{r}_1,\mathbf{r}_2) \frac{\mathbf{r}_3 - \mathbf{r}_1}{|\mathbf{r}_3 - \mathbf{r}_1|^{n+1}} \cdot \mathbf{e}_{12},
     \end{equation}
where $R_0$ is a normalization scale constant. This parametrization is equivalent to the standard gravity expression for $n=2$; however due to its linear mass dependence, it is not equivalent to MOND in the $n=1$ case. It can be shown that this expression reduces to 

\begin{eqnarray}
    g(\vec x) = \frac{3}{4x^2}\frac{H^2a\,\Omega_m R_0^{n-2}}{1+\xi_{gg}(x)}\int zdz\,\xi_g(z)F_n(x,z),
\end{eqnarray}
where \begin{multline}
    F_n(x,z) = (x^2-z^2)\frac{(x+z)^{1-n}-|x-z|^{1-n}}{1-n}+\\ \frac{(x+z)^{3-n}-|x-z|^{3-n}}{3-n},
\end{multline}
for $n>1$, and for n=1 tends to 
\begin{equation}
    F_1(x,z) = (x^2-z^2)\log\frac{x+z}{|x-z|} + 2xz.
\end{equation}

\begin{figure}
    \includegraphics[width=0.44\textwidth]{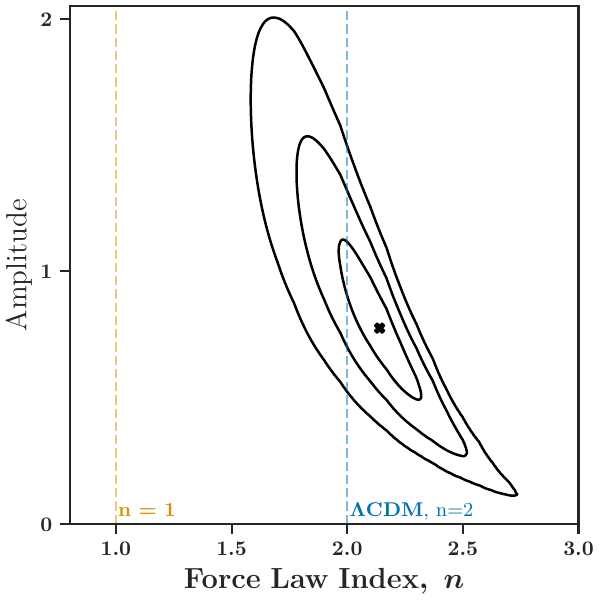}
    \caption{Likelihood for the alternative gravitational model described by Eq. \ref{eq:altgraveq} using a length scale $R_0=250\,\rm Mpc$. The exponent is constrained to $n>1.8$ at 95\% confidence and the case $n=1$ is disfavored at $6\sigma$. Note that the $n=1$ case does not correspond to MOND in this model.}
    \label{fig:altgravtwofiftyMpcNormalization}

\end{figure}

Using the same procedure presented above, this parametrization constrains the exponent to $n>1.8$ at 95\% confidence, and disfavors the case $n=1$ at $>6\sigma$ as shown in Figure \ref{fig:altgravtwofiftyMpcNormalization}. Note that due to its different mass dependence, this model does not directly correspond to MOND for any parameter $n$.

\bibliographystyle{act_titles2_unsrt}
\bibliography{bibliography}

\nocite{Peebles1976, Davis1977, Peebles1976, Juszkiewicz1999, Sheth2001, Juszkiewicz1999, Sheth2001, Seto1999, Peebles1980, Milgrom1983, Bekenstein1984, Sanders2002, Kosowsky2010, Famaey2012, McGaugh2004, McGaugh2005, McGaugh2012, Bekenstein2004, Abbott2017, LIGOGRTests2019, Boran2018, Nipoti2008, McGaugh2010, Safarzadeh2021, Sanders1999, Sanders2003, Tian2020, Dodelson2011, Pardo2020, skordis_PhysRevLett.127.161302, Bekenstein2004, Milgrom1983, Famaey2012, McGaugh2013, Bekenstein2006, Bekenstein2006, Milgrom1986, Slosar2008, Cabass:2022wjy, Cabass:2022ymb, Slosar2008, Cabass:2022wjy, Cabass:2022ymb}

\end{document}